\documentstyle[preprint,aps]{revtex}
\begin{document}
\draft
\preprint{\vbox{Submitted to Physical Review {\bf C}\hfill
        UC/NPL-1116}}
\tolerance = 10000
\hfuzz=5pt
\tighten
\begin{title}
{Isospin-breaking corrections to nucleon electroweak \\
form factors in the constituent quark model}
\end{title}
\author{V.~Dmitra\v sinovi\' c and S.J. Pollock}
\address{
   Nuclear Physics Laboratory, Physics Department,\\
   University of Colorado, P.O.Box 446,
   Boulder, CO 80309-0446
\\Email addresses:  dmitra@godot.colorado.edu, pollock@lucky.colorado.edu}
\maketitle
\begin{abstract}

We examine isospin breaking in the nucleon wave functions due to the
$u - d$ quark mass difference and the Coulomb interaction among the
quarks, and their consequences on the nucleon electroweak form factors
in a nonrelativistic constituent quark model.  The mechanically induced
isospin breaking in the nucleon wave functions and electroweak form
factors are exactly evaluated in this model. We calculate the
electromagnetically induced isospin admixtures by using first-order
perturbation theory, including the lowest-lying resonance with nucleon
quantum numbers but isospin 3/2.  We find a small ($\leq 1\%$), but
finite correction to the anomalous magnetic moments of the nucleon
stemming almost entirely from the quark mass difference, while the
static nucleon axial coupling remains uncorrected.  Corrections of the
same order of magnitude appear in charge, magnetic, and axial radii of
the nucleon. The correction to the charge radius in this model is
primarily isoscalar, and may be of some significance for the extraction
of the strangeness radius from e.g. elastic forward angle parity
violating electron-proton asymmetries, or elastic ${}^4He({\vec e},e')$
experiments.

\end{abstract}
\newpage

\section{Introduction}

Parity-violating electroweak lepton-nucleon
scattering such as $N({\vec e}, e')N,~N({\nu}, {\nu}')N$ is of
particular interest to subnucleon physics because it is sensitive to
the Standard Model fundamental coupling constants $\alpha ,
\sin^{2}\theta_{W}$ and various polar- and axial-vector current form
factors of the nucleon \cite{rev94}.  The standard analysis, that led
to the current interest in these processes, is based on the following
assumptions:
(i) Lorentz + translational invariance,
(ii) Standard Model of lepton and quark electroweak interactions,
(iii) one boson exchange approximation,
(iv) significance of only $u$ and $d$ quarks in the nucleon states,
(v) good parity of the nucleon states, and
(vi) good isospin of the nucleon states.
In view of pending $N({\vec e}, e')N$ experiments \cite{prop91}, and of
pending and completed elastic $N({\nu}, {\nu}')N$ \cite{bnl92}
measurements, it seems worthwhile to explore the corrections due to the
relaxation of some, or all, of these premises. So far, assumptions
(ii-v) have been relaxed and their consequences explored
\cite{rev94,vd92}. In this note we will relax assumption (vi) in a
specific quark model of the nucleon and look at its effects on
electroweak form factors of the nucleon. The results are of immediate
relevance to the interpretation of the isoscalar axial coupling as
measured e.g. in elastic neutrino-nucleon scattering
\cite{bnl92,hor93,gar93}.  They will also be relevant for isoscalar
vector couplings extracted in future neutrino and parity violation
experiments.  This is because intrinsic isospin breaking, although
expected to be small, modifies observables in much the same way as
nonzero strangeness content of the nucleon does.

A common present-day picture of the nucleon \cite{bhad88,yopr88,wal95} is one
of the ground state of three quarks bound by gluon exchange according to
quantum chromodynamics (QCD).  Given the intractable nature of this
strong coupling few-body problem, we resort to a nonrelativistic
constituent quark model with harmonic oscillator confining quark-quark
interactions, chosen for its simplicity. The isospin breaking (IB) in
our model is due to the following two sources: quark mass differences, and
electroweak interactions, which is essentially the same as in QCD \cite{isg80}.
Since both of these corrections are expected to be small relative to the
isospin-conserving components, we will evaluate each one in its own
right without considering the cross-terms. The results of our analysis
will justify this assumption {\it ex post facto}. Furthermore, we will
confine ourselves to the investigation of electroweak interactions {\it
among quarks in the nucleon} and neglect their self-interactions.
The results will turn out to be particularly sensitive to the shift in the
nucleon's center of mass, which is faithfully reflected in this model.

In this paper, we begin (Section II) with a general discussion of
isospin breaking and the connection to electroweak form factors. The
purpose is to formalize and clarify the extent to which e.g. the
strangeness content of the nucleon can be distinguished from IB
modifications in a model independent way. In section III, we describe
the specific quark model we use, and outline the calculation of IB
effects due to electromagnetic quark-quark interactions, and quark mass
differences.  In section IV, we present the results of these
calculations and discuss them.  We find that some IB effects are
significantly larger than one might naively expect. Only one of the
corrections (the weak neutral current charge radius) we have found
appears to present possible qualitative problems for extracting
non-trivial strangeness content, if that should turn out to be present.
The calculations presented in sections III and IV are model dependent,
and are manifestly just a part of a larger set of corrections. We
conclude section IV with a brief discussion of future directions to be
pursued.  In section V, we summarize our results and draw our
conclusions.

\section{Model-independent formalism}

One can define isospin at two different levels: (a) at the level
of quarks, where the Standard Model is constructed; and (b) at the hadron
(here, nucleon) level, where all of experimental data originate.
We will need both levels to define IB in our formalism. That fact limits
the applicability of the present formalism to {\em quark} models of the
nucleon and leaves out nucleon models with purely hadronic degrees of freedom
such as Skyrme's.
Attempts to define isospin-breaking admixtures to the nucleon wave functions
exclusively at level (b)
can easily lead to tautologies and circular arguments. In order to avoid
that we will first review the consequences of exact isospin, and then
introduce notation which should  clearly denote the origin and
character of the various contributions to the total (physical) form factors.

In the isospin-symmetric limit the isospin transformation properties of the
operators are identical to those of the matrix elements, as a direct
consequence of the Wigner-Eckart theorem.
The main consequence of isospin breaking in the nucleon wave functions
is the loss of these, previously exact, isospin transformation properties of
the electroweak current matrix elements as compared with those of the current
operators themselves.
A simple example will best clarify this statement. The third
component of the isovector axial current in the ``nuclear domain", where only
$u,d$ quarks are kept, reads \footnote{Note that our conventions on the
definitions of electroweak currents differ slightly from ref. \cite{rev94}.
We adopt the Bjorken and Drell (BD) \cite{bd65} metric and conventions, with
the exception of the normalization of the Pauli form factor which we take to be
$F_{2}(0) = \kappa$.}
\begin{equation}
J_{\mu 5}^{z} \ =\  {\bar q} {{\mbox{\boldmath$\tau$}}^{z} \over 2}
\gamma_{\mu} \gamma_{5} q ~~~.
\end{equation}
When sandwiched between two nucleon states $|N \rangle$ with ``impure"
isospin, this does {\it not } yield just $+ {1\over 2} F_{A}$ for the
proton and $- {1 \over 2} F_{A}$ for the neutron, but rather it induces
a small but finite effective isoscalar axial form factor.  To see the
relevance of these comments, we turn to the weak neutral current (NC)
${\cal J}_{\mu}^{\rm NC} = J_{\mu}^{\rm NC} - J_{\mu 5}^{\rm NC}$. The
axial neutral current, in the presence of three flavors, is
\begin{equation}
J_{\mu 5}^{\rm NC}\ =\ {1\over2}\left(
	\bar u\gamma_\mu\gamma_5 u  -
	\bar d \gamma_\mu\gamma_5 u -
	\bar s \gamma_\mu\gamma_5 s \right)~~~.
\end{equation}
$J_{\mu 5}^{\rm NC}$ consists of the nuclear domain isovector current
and a strange quark part. It is these ``strange" (hidden
strangeness-induced) terms that are a topic of considerable recent
interest.  They would {\it also} yield an effective isoscalar axial
form factor.  The hidden strangeness and intrinsic isospin breaking
effects cannot be separated by experiment, thus any nontrivial isospin
breaking could potentially affect a determination of the strange form
factors.

Similar, but not identical comments hold for the charge and magnetic
form factors, because both the isoscalar and isovector current
operators enter the EM current. In the following we will define the
effects of such isospin-breaking in the nucleon wave functions on
electroweak form factors.  We define the axial-vector current form
factors for real nucleon states as follows:
\begin{equation}
\langle N(p')|{1\over2}(\bar u \gamma_{\mu} \gamma_{5} u - \bar d
\gamma_{\mu} \gamma_{5}  d) |N(p) \rangle  \equiv
 \bar u(p') \left[ {1\over2}({}^{u - d}F_{A}^{p + n} \pm
{}^{u - d}F_{A}^{p - n}) \gamma_{\mu} \gamma_{5}  \right] u(p)~~~, \
\end{equation}
where $\pm = \langle N |\mbox{\boldmath$\tau$}_{N}^{\rm z} | N \rangle$
for proton or neutron matrix elements, respectively, and
$q^{2} = q_{0}^{2} - {\bf q}^{2}$ is the invariant momentum transfer.
The notation for the form factor superscripts is as
follows: the upper left-hand
index refers to the quark operator on the left-hand side side of the equation.
The upper right-hand index refers to the {\it nucleon} isospin operator.
In the above case, the quark operator is $u-d$ and hence quark-isovector. It
induces a large nucleon-isovector ${}^{u - d}F_{A}^{p - n}$ axial form factor
and a small nucleon-isoscalar axial form factor ${}^{u - d}F_{A}^{p + n}$
correction.
We also define the ``strange" form factors $S_{1,2,A}$ as follows
\begin{eqnarray}
\langle N(p')|(\bar s \gamma_{\mu} \gamma_{5} s ) |N(p) \rangle  &\equiv&
 \bar u(p') \left[ S_{A}(q^2) \gamma_{\mu} \gamma_{5}  \right] u(p), \\
\langle N(p')|(\bar s \gamma_{\mu}  s ) |N(p) \rangle  &\equiv&
 \bar u(p') \left[ S_{1}(q^2) \gamma_{\mu}
+{i \sigma_{\mu\nu}q^\nu\over 2 M_N} S_{2}(q^2)
 \right] u(p)~~~,
\end{eqnarray}
which are all isoscalar to first approximation.
We ignore here any (isovector) IB {\it corrections} to the strange
form factors of the nucleon, since the strange form factors themselves
can be viewed as a consequence of flavor SU(3) symmetry breaking admixtures
and hence are expected to be small relative to the $u, d$ induced ones,
thus making the IB corrections to them ``doubly small" i.e. a second order
effect.
$S_1(0) = 0$ follows from the fact that the
nucleon has zero total strangeness.

With our conventions, the EM current operator reads
\begin{equation}
J_\mu ^{\rm EM} \ =\
{1\over2} (
\bar u\gamma_\mu u - \bar d \gamma_\mu d) +
{1\over6} ( \bar u\gamma_\mu u + \bar d \gamma_\mu d
- 2 \bar s \gamma_\mu s)~~~.
\end{equation}
The polar-vector part of the weak neutral current operator is
\begin{equation}
J_\mu ^{\rm NC}\ =\
{1\over2} ( \bar u\gamma_\mu u - \bar d \gamma_\mu d) -
{1\over2}  \bar s \gamma_\mu s - 2 \sin^2\theta_W J_\mu ^{\rm EM}~~~,
\end{equation}
where $\theta_{W}$ is the weak mixing angle.
We define polar-vector form factors for real nucleon states
as follows:
\begin{eqnarray}
\langle N(p')| {1\over2}(\bar u \gamma_\mu u && - \bar d \gamma_\mu d) |N(p)
\rangle  \equiv \nonumber \\
\bar u(p') && \left[ {1\over2}({}^{u - d}F_{1}^{p + n} \pm
{}^{u - d}F_{1}^{p - n}) \gamma_{\mu}
+  {1\over2}({}^{u - d}F_2^{p + n} \pm {}^{u - d}F_2^{p - n})
{i \sigma_{\mu\nu}q^\nu \over 2 M_{N}} \right] u(p)~~~. \\
\langle N(p') | {1\over6}(\bar u \gamma_\mu u && +
\bar d \gamma_\mu d) | N(p) \rangle \equiv  \nonumber \\
\bar u(p') &&
\left[ {1\over2}({}^{u + d}F_1^{p + n} \pm {}^{u + d} F_1^{p - n})
\gamma_\mu +  {1 \over2}({}^{u + d} F_2^{p + n} \pm {}^{u + d}F_2^{p - n})
{i \sigma_{\mu\nu}q^\nu \over 2 M_{N}}  \right] u(p)~~~. \
\end{eqnarray}

In this notation, if isospin were a good symmetry of the nucleon
states, and there were no strangeness content, then ${}^{u - d}
F^{p+n}(q^{2}) = {}^{u + d}F^{p-n}(q^{2}) = 0$ (by the Wigner-Eckart
theorem), and ${}^{u + d}F^{p+n}(q^{2})$ would be the usual nucleon
isoscalar form factor, ${}^{u - d}F^{p - n}(q^{2})$ would be the usual
nucleon isovector form factor.\footnote{Note that we have absorbed a
factor of $1/2$ into the definition of the quark isovector, ``u-d'',
superscript, and a factor of $1/6$ into the quark isoscalar, ``u+d'',
notation.} With the above definitions, we can write weak matrix
elements in terms of the above electromagnetic and the strange ones
with some, as yet unknown,  nucleon IB corrections.  For example,
\begin{equation}
F_{i,N}^{\rm NC}\  =\  {1\over2} ({}^{u - d}F_{i}^{p + n}\pm {}^{u - d}
F_{i}^{p - n}) - 2 \sin^2 \theta_W F_{i,N}^{EM} - {1\over2} S_i~~,
\end{equation}
where $i = 1,2$, i.e. for polar-vector neutral current form factors.
Similarly, we have for the axial-vector NC form factors
\begin{equation}
F_{A,N}^{\rm NC}\  =\  {1\over2} ({}^{u - d}F_{A}^{p + n}
\pm {}^{u - d}F_A^{p-n}) - {1\over2} S_A~~.
\end{equation}
Taking the difference between the proton and the neutron polar-vector NC
form factors, we find
\begin{equation}
F_{i,p}^{\rm NC} - F_{i,n}^{\rm NC} = (1-2\sin^2 \theta_W)
				(F_{i,p}^{\rm EM}- F_{i,n}^{\rm EM})
	-{}^{u + d}F_{i}^{p - n}~~.
\label{isovector}
\end{equation}
Here the isospin breaking form factor ${}^{u+d}F_i^{p-n}$ appears as a
correction to the usual isovector form factor relation, and strangeness
content cancels out.
In principle, this means that
the IB isovector correction ${}^{u + d}F_{i}^{p - n}$
is measurable to this order in
perturbation theory, given a complete set of experiments, i.e. if we
had both the proton and the neutron experiments.
Similarly, we find for the isoscalar part
\begin{equation}
F_{i,p}^{\rm NC} + F_{i,n}^{\rm NC} = -2\sin^2\theta_W
                                (F_{i,p}^{\rm EM} + F_{i,n}^{\rm EM})
	-S_i
        +{}^{u - d}F_{i}^{p + n}~~.
\label{isoscalar}
\end{equation}
The final term again represents isospin breaking effects.
For the axial form factors we find
\begin{eqnarray}
F_{A,p}^{\rm NC} - F_{A,n}^{\rm NC} &\ = &\  {}^{u - d} F_{A}^{p - n}~~,\\
F_{A,p}^{\rm NC} + F_{A,n}^{\rm NC} &\ = &\
        {}^{u - d}F_{A}^{p + n} - S_{A}~~.
\label{iso_a}
\end{eqnarray}
As discussed above, the isospin correction terms in
Eqs.~(\ref{isoscalar}) and (\ref{iso_a}) enter exactly like the strange
form factors do.  All of these IB corrections are expected to be
extremely small, of course, and will be estimated in this paper.

Finally, there are some constraints on the IB admixture-induced form
factors at zero momentum transfer stemming from exact symmetries.
Gauge invariance says that the electric charge of the proton must be
unity, so $F_{1,p}(0) = 1$, and $F_{1,n}(0) = 0$.  In fact, both
${}^{u + d}F_1^{p - n}(0) $ and ${}^{u - d}F_1^{p + n}(0) $
vanish identically at zero momentum transfer to leading order in IB
interactions; the first relation is due to the Ademollo-Gatto
\cite{ag64} theorem, the second due to baryon number conservation.
The latter result follows from the orthogonality of excited states
used in the first-order perturbation theory and the fact that the
charge form factor reduces to the norm of the ground state, at zero momentum
transfer.
This statement does not apply to the magnetic or axial form factors at any
$q^2$, nor to the electric one at nonvanishing momentum transfers. Such
correction terms will be typically of order $\alpha$.  In the next
section, we  calculate, in a simple constituent quark model, the IB
corrections to the Walecka-Sachs \cite{wal59} form factors
${}^{u + d}G_{i}^{p - n}(q^{2})$
and ${}^{u - d}G_{i}^{p+n}(q^{2})$, $i = E, M$,
which are related to the Dirac and Pauli form factors $F_{1,2}$ by
\begin{eqnarray}
G_{E}(q^2) &=&  F_{1}(q^2) + {q^2 \over 4 M_{N}^{2}} F_{2}(q^2) \nonumber \\
G_{M}(q^2) &=&  F_{1}(q^2) + F_{2}(q^2) ~~~. \
\end{eqnarray}

\section{The Model}

We use the constituent quark model \cite{bhad88,yopr88} to calculate both of
the above mentioned
isospin-breaking corrections and use the harmonic oscillator Hamiltonian for
that purpose.
This three-body problem can be reduced to two uncoupled harmonic oscillators
by application of the (equal mass) Jacobi three-body coordinates
\begin{mathletters}
\begin{eqnarray}
\mbox{\boldmath$\rho$} &=& {1 \over{\sqrt{2}}}
\left({\bf r}_{1} - {\bf r}_{2}\right)
				\label{eqiia}	 \\
\mbox{\boldmath$\lambda$} &=& {1 \over{\sqrt{6}}}
\left({\bf r}_{1} + {\bf r}_{2} - 2 {\bf r}_{3}\right)
\label{eqiib}\\
{\bf R} &=& {1 \over 3} \left( {\bf r}_{1} +  {\bf r}_{2} +
{\bf r}_{3}\right)  ~.
\label{eqiic}\
\end{eqnarray}
\end{mathletters}
Then, the Hamiltonian consists of two independent harmonic oscillators with
equal mass $m$ and a freely moving  center of momentum with mass $M = 3 m$,
\begin{eqnarray}
H = {{\bf P}^{2} \over 2 M} + {{\bf p}_{\rho}^{2} \over 2 m} +
{{\bf p}_{\lambda}^{2} \over 2 m} + {3 k \over 2}
\left({\mbox{\boldmath$\rho$}}^{2} + {\mbox{\boldmath$\lambda$}}^{2}\right)~~.
\end{eqnarray}
Solutions to the Schr\" odinger equation with this Hamiltonian are well
known and have been tabulated in \cite{bhad88,yopr88,jmr92} for low-lying
nucleon resonances.  A simple confining interaction such as the
harmonic oscillator leads to spatial wave functions that are Gaussians,
in the case of the ground state, or that decay as Gaussians at large
distances, for any other state of the system. That, in turn, leads to
Gaussian EM and axial form factors which fall off far too rapidly at
high values of momentum transfer compared with experiment. This model
is clearly rather simple, but should be adequate for the purposes
of identifying qualitative features, and making first estimates of
small IB effects. In this spirit we have neglected the
strong-hyperfine-interaction interference with the IB terms in the
Hamiltonian \cite{isg80,jmr92}. The former is an important part of the
extended constituent quark model \cite{ik79}, and its effects have been
evaluated for simple IB observables such as the hadron mass differences
\cite{isg80,jmr92}. But, in our case it complicates the
evaluation of the observables to such an extent that we relegate its
inclusion to the future.

\subsection{Mechanical Corrections}

The notion of quark mass is an ill-defined one: free quarks have not
been observed. Traditionally one distinguishes between two kinds of
quark masses: (1) the current quark mass $m_{q}$ which is one that the
quarks would have in the absence of all strong interactions; (2) the
constituent quark mass $m_{Q}$ which is the measure of inertia of a
quark moving within hadrons.  There is of course considerable
difficulty in precisely pinning down the values of the current quark
masses \cite{pdg94}, and even more in connecting these to constituent
quark masses.  Weinberg \cite{wein77} has used chiral symmetry and
empirical meson masses to argue that $m_{u} \simeq 5 {\rm MeV}; m_{d}
\simeq 9 {\rm MeV}; m_{s} \simeq 140 {\rm MeV}$ and $m_{Q} = m_{q} +
{\rm const}$, where ${\rm const} \simeq 330 {\rm MeV}$. This, in turn,
implies
\begin{equation}
\Delta m_{q} \equiv m_{u} - m_{d} = \Delta m_{Q} \equiv m_{U} - m_{D}
\simeq - 4 {\rm MeV}, \label{qmasses}
\end{equation}
and those are the values that we use\footnote{Isgur \cite{isg80} finds
$\Delta m_{q} = - 6 {\rm MeV}$ in his extended version
of the present model that includes the strong hyperfine interaction effects.}.
Consequently, we expect the
leading isospin breaking corrections due to explicit quark mass
differences (which we call ``mechanical corrections'') to be of
${\cal O}({\Delta m_{Q} / m_{Q}})$, where
$ {\Delta m_{Q} / m_{Q}}\simeq - 1 /85$
i.e.  rather small. They can be exactly evaluated if one assumes
harmonic oscillator confining quark-quark interactions with
{\it unequal} masses,
\begin{eqnarray}
H = \sum_{i= 1}^{3}{{\bf p}_{i}^{2} \over 2 m_{i}} + {k \over 2}
\sum_{i < j}^{3}({\bf r}_{i} - {\bf r}_{j})^{2} ,
\end{eqnarray}
where $m = m_{1} = m_{2} \neq m_{3} = m^{'}$.
The first two ($\mbox{\boldmath$\rho$}$ and $\mbox{\boldmath$\lambda$}$) of
the three-body Jacobi coordinates
in this case are the same as Eqs.~(\ref{eqiia}) and (\ref{eqiib}), but
now carry quark number indices e.g.
\begin{mathletters}
\begin{eqnarray}
\mbox{\boldmath$\rho$}_{3} &=& {1 \over{\sqrt{2}}}
\left({\bf r}_{1} - {\bf r}_{2}\right)
				\label{ja}	 \\
\mbox{\boldmath$\lambda$}_{3} &=& {1 \over{\sqrt{6}}}
\left({\bf r}_{1} + {\bf r}_{2} - 2 {\bf r}_{3}\right)~~,
\label{jb}\
\end{eqnarray}
\end{mathletters}
and the center of mass is shifted to
\begin{equation}
{\bf R} = {1 \over{(2 m + m^{'})}} \left(m {\bf r}_{1} + m {\bf r}_{2} +
m^{'} {\bf r}_{3}\right)~.
\label{eqvi}\
\end{equation}
This allows a separation of variables and an {\it exact} solution to
the problem in terms of two harmonic oscillator wave functions with two
{\it different} masses,
\begin{equation}
m_{\lambda} = {3 m m' \over{2 m + m'}};
{}~m_{\rho} = m,
\label{masses}
\end{equation}
and a freely moving  center of momentum with mass $M_{N} = 2 m +
m^{'}$
(where $m=m_u,\ m'=m_d$ in the proton, $m=m_d,\ m'=m_u$ in the neutron):
\begin{equation}
H = {{\bf P}^{2} \over 2 M_{N}} + {{\bf p}_{\rho}^{2} \over 2 m_{\rho}} +
{{\bf p}_{\lambda}^{2} \over 2 m_{\lambda}} + {3 k \over 2}
\left({ \mbox{\boldmath$\rho$}}^{2} + {\mbox{\boldmath$\lambda$}}^{2}
\right) . \label{eqvii}
\end{equation}
The total wave function factors into the CM plane wave solution and an
internal motion wave function
\begin{equation}
|~\Psi_{N}({\bf P}_{i}) \rangle = \left({1 \over{3}}\right)^{3/4}
\exp \left(i ({\bf P}_{i} \cdot {\bf R} - E_{i} t)\right)
|N (940) \rangle~~, \label{cmwf}
\end{equation}
where
$E_{N}^{i} = {{\bf P}_{i}^{2} \over 2 M_{N}} +
{3 \over 2}\left(\omega_{\rho} + \omega_{\lambda}\right)$,
$\omega \equiv \sqrt{3 k/m}$, and the factor
$\left({1 \over{3}}\right)^{3/4}$ is the
square root of the inverse Jacobian for the above three-body Jacobi
coordinates. The spring constant $k$ is an adjustable parameter of
this model.

For spin- and isospin-independent quark potentials the complete internal
wave function factorizes into a product of the spin, isospin, and spatial
wave functions. Once the isospin-dependent terms are introduced
into the Hamiltonian, however, the factorization property breaks down and
the spatial and isospin parts of the internal wave function become entangled
\begin{equation}
^{2}8:~~|N (940) \rangle = {1 \over \sqrt2}
\left[ \chi^{\rho} \Phi^{\rho} +
\chi^{\lambda} \Phi^{\lambda} \right]
\label{wave}
\end{equation}
The spin parts are given in a standard notation:
\begin{eqnarray}
\chi_{\uparrow}^{\rho} &=& {1 \over{\sqrt 2}}
\left(\alpha \beta - \beta \alpha \right)\alpha \nonumber  \\
\chi_{\uparrow}^{\lambda} &=& {1 \over{\sqrt 6}}
\left(2 \alpha \alpha \beta  - \alpha \beta \alpha - \beta \alpha \alpha
\right),~~\
\label{spinwave}
\end{eqnarray}
while the spatial-isospin parts for the proton are given by
\begin{mathletters}
\begin{eqnarray}
\Phi_{p}^{\rho} &=& {1 \over{\sqrt 2}}
\left(\psi_{0}(udu) - \psi_{0}(duu) \right) \nonumber  \\
\Phi_{n}^{\rho} &=& {1 \over{\sqrt 2}}
\left(\psi_{0}(dud) - \psi_{0}(udd) \right)  \nonumber \\
\Phi_{p}^{\lambda} &=& {1 \over{\sqrt 6}}
\left(2 \psi_{0}(uud) - \psi_{0}(udu) - \psi_{0}(duu)\right) \nonumber \\
\Phi_{n}^{\lambda} &=& {1 \over{\sqrt 6}}
\left(2 \psi_{0}(ddu) - \psi_{0}(dud) - \psi_{0}(udd)\right)~~,\
\label{isowave}
\end{eqnarray}
\end{mathletters}
where
\begin{eqnarray}
\psi_{0}(udu)
&=& \left({m_{\rho} \omega_{\rho} \over{\pi}}\right)^{3/4}
\left({m_{\lambda} \omega_{\lambda} \over{\pi}}\right)^{3/4}
\exp\left(- {1 \over 2}
\left({{\mbox{\boldmath$\rho$}}_{2}^{2} \over R_{\rho}^{2}} +
{{\mbox{\boldmath$\lambda$}}_{2}^{2} \over R_{\lambda}^{2}}\right) \right)
\otimes udu  \label{ibwf} \\
&\simeq&
\left({m_{u} \omega_{u} \over{\pi}}\right)^{1/2}
\left({m_{d} \omega_{d} \over{\pi}}\right)^{1/2}
\left({m_{u} \omega_{u} \over{\pi}}\right)^{1/2} \otimes udu
\nonumber  \\
&\times& \exp\left(- {1 \over 2}
\left(\omega_{u} m_{u} ({\bf r}_{1} - {\bf R})^{2} +
\omega_{d} m_{d} ({\bf r}_{2} - {\bf R})^{2} +
\omega_{u} m_{u} ({\bf r}_{3} - {\bf R})^{2}\right) \right)
\nonumber ~~, \
\end{eqnarray}
the second line of Eq. (\ref{ibwf}) is meant in Faiman and Hendry's
\cite{fh68} symbolic sense only and is not to be used in actual calculations.
Here $R_{\alpha}^{-2} = m_{\alpha} \omega_{\alpha} = \sqrt{3 k m_\alpha},
{}~~\alpha = \rho, \lambda$.
The corresponding neutron expression is found by switching $u$ with
$d$ everywhere.
The simpler isospin-{\it symmetric} wave function \cite{bhad88,yopr88} can
be read off directly by setting
$m_u = m_d \equiv m$. It factorizes into three parts as expected. The
complete wavefunction is crucial to our subsequent work, however, as
straightforward use of the standard factorized nucleon wave functions
fails to correctly treat isospin breaking.

We are primarily interested in the elastic charge and magnetic form
factors $G_{\rm E}({\bf q}^{2})$, $G_{\rm M}({\bf q}^{2})$ defined as the
Fourier transforms of the charge and current densities
\begin{eqnarray}
\intop\limits d {\bf R} && \langle \Psi_{N}({\bf P}^{'})|
		{\bf J}_{\rm EM}({\bf R})|~\Psi_{N}({\bf P}) \rangle
 \exp (i {\bf q} \cdot {\bf R})\ =  \nonumber\\
&& = (2 \pi)^{3} \delta ({\bf P}^{'} - {\bf q} - {\bf P}) e \left[
\left({{\bf P} + {\bf P}^{'} \over 2 M_{N}}\right)
\langle {\it 1} \rangle_{N} G_{\rm E}({\bf q}^{2})
+ i  \left({\langle{ \mbox{\boldmath$\Sigma$}} \rangle_{N}
\times {\bf q}
\over 2 M_{N}} \right) G_{\rm M}({\bf q}^{2}) \right] \
\label{emc}
\end{eqnarray}
in the nonrelativistic limit,
where $|\Psi_{N} \rangle $ is the ground state of the exact Hamiltonian
Eq.~(\ref{eqvii}),
${\bf P}$ and ${\bf P}^{'}$ are the CM momenta of the initial and final state
nucleons, respectively, and ${\bf q}={\bf P}^{'}-{\bf P}$ is the three-momentum
transfer. We work in the Breit frame defined by
${\bf q} = 2 {\bf P}^{'} = - 2 {\bf P}$, which ensures that $E = E^{'}$.
In the Eq. (\ref{emc}) above, $\bf R$ is the photon position vector and
{\em not} the CM position vector,
$\mbox{\boldmath$\Sigma$}$ and ${\it 1}$ are the Pauli and
unit matrices, respectively, operating in the nucleon spin space and
$\langle{ \mbox{\boldmath$\Sigma$}} \rangle_{N}$,
$\langle {\it 1} \rangle_{N}$ are their matrix elements taken between nucleon
spinors.
The quark EM current density operator is the nonrelativistic reduction of the
Dirac fermion current
\begin{eqnarray}
{\bf J}_{\rm EM}({\bf R}) &=& \sum_{i=1}^{3}  {e_{i} \over 2 m_{i}}
\left( - i\left\{ {\bf \nabla}_{{\bf r}_{i}},
\delta ({\bf r}_{i} - {\bf R})  \right\} +
 { \mbox{\boldmath$\sigma$}}_{i} \times  \left[{\bf \nabla}_{\bf R},
\delta ({\bf r}_{i} - {\bf R})
 \right] \right) , \
\end{eqnarray}
where $e_{i} = {1 \over 2} ({1 \over 3} + {\mbox{\boldmath$\tau$}}_{i}^{z})$
is the quark electric charge operator. Similarly, for the axial current in the
isospin-symmetric limit, we have
\begin{eqnarray}
\intop\limits d {\bf R} && \langle \Psi_{N}({\bf P}^{'})|
		{\bf J}_{\rm A}^{a}({\bf R})|~\Psi_{N}({\bf P}) \rangle
\exp (i {\bf q} \cdot {\bf R}) \nonumber \\
&& = (2 \pi)^{3} \delta ({\bf P}^{'} - {\bf q} - {\bf P})
\langle {\mbox{\boldmath$\Sigma$}} \rangle_{N} \
\langle {1 \over 2}\mbox{\boldmath$\tau$}_{N}^{a} \rangle_{N}
F_{\rm A}({\bf q}^{2})~~ . \
\end{eqnarray}
Here, $\mbox{\boldmath$\tau$}_{N}$ are the isospin matrices for the nucleons,
and the quark axial current reads
\begin{eqnarray}
{\bf J}_{\rm A}^{a}({\bf R}) &=& \sum_{i=1}^{3}
 {{\mbox{\boldmath$\tau$}}_{i}^{a}
\over 2} { \mbox{\boldmath$\sigma$}}_{i}\delta ({\bf r}_{i} - {\bf R})~~ . \
\end{eqnarray}
Thus we have made yet another assumption: our constituent quark
electroweak interactions are identical to those of the corresponding
current quarks. Specifically, this means that we assume vanishing
anomalous magnetic moments and no form factors for constituent quarks.
For a discussion of this point see section IV.C.

Despite the exact solvability of the Schr\" odinger equation with the
Hamiltonian Eq.~(\ref{eqvii}),
we will expand the exact form factors in powers of ${\bf q}^2$.
The reason for this is that a simple confining interaction
such as the harmonic oscillator leads to Gaussian spatial wave
functions, which in turn lead to an unrealistically rapid form factor
fall-off at high ${\bf q}^2$.
Therefore, it only makes sense to talk about the leading terms
in the expansion in powers of momentum transfer.

\subsection{Electromagnetic Corrections}

The electroweak interactions among quarks are readily divided into
parity-conserving electromagnetic (EM) terms that are of
${\cal{O}}(\alpha)$, where $\alpha \simeq 1/137$ is the fine-structure
constant, and the parity-violating weak interactions that are of
${\cal{O}}(G_{\rm F})$ where $G_{F} \simeq (1.023 \pm 0.002) \times
10^{-5} M_{p}^{2}$ is the Fermi weak coupling constant. The latter
terms lead to parity-admixture corrections that have already been
analyzed in Ref.\cite{vd92}. Since we are working in the
nonrelativistic approximation, we expand the invariant M{\o}ller
operator in powers of momentum over mass, which leads to the Coulomb
interaction as the leading EM term
\begin{eqnarray}
V_{\rm EM} = \sum_{i < j}^{3}{e_{i} e_{j} \over{4 \pi
|{\bf r}_{i} - {\bf r}_{j}|}} .
\end{eqnarray}
The hyperfine and the spin-orbit coupling interactions are ``higher-order"
corrections in ${p \over m_{Q}}$.
The Coulomb interaction $V_{\rm EM}$ induces a small ${\cal{O}}(\alpha)$
abnormal isospin admixture in
the nuclear wave function. To determine this admixture we use the first-order
(Rayleigh-Schr\" odinger) perturbation theory in the perturbing Coulomb
potential, as applied to the Hamiltonian
\begin{eqnarray}
H = H_{0} + V_{\rm EM} ~~.
\end{eqnarray}
The ground state of the nucleon $| \Psi_{0} \rangle$, to ${\cal{O}}({\alpha})$,
is given by
\begin{eqnarray}
| \Psi_{0} \rangle = |\Phi_{0} \rangle  + \sum_{n \neq 0} |\Phi_{n} \rangle
{ \langle \Phi_{n}|V_{\rm EM} |\Phi_{0} \rangle
\over{E_{0} - E_{n}}} + {\cal{O}}({\alpha}^2),
\end{eqnarray}
where $|\Phi_{n}\rangle$ are the {\em exact} eigenstates of the
isospin-symmetric Hamiltonian $H_{0}$: $H_{0} |\Phi_{n} \rangle = E_{n}
|\Phi_{n} \rangle$.  This involves evaluating the isospin-breaking
admixture to the nucleon wave function defined by the admixture
coefficients $\varepsilon_{n}$:
\begin{equation}
\varepsilon_{n} = { \langle \Phi_{n}|V_{\rm EM} |\Phi_{0} \rangle
\over{E_{0} - E_{n}}} \label{eps}
\end{equation}
The sum extends over {\it all} $n$, i.e., over infinitely many excited
states of the nucleon.

The isospin-admixtures in the wave function
generate isospin breaking contributions to the elastic
nucleon electroweak current matrix elements. These
are determined by using the first-order perturbation theory
parameters $\varepsilon_{n}$ and the vector and axial current
transition matrix elements between the nucleon and its excited states
calculated in the non-relativistic impulse approximation.  The
definition of the isospin-violating corrections to the non-relativistic
current matrix elements to ${\cal{O}}(\alpha)$ is thus
\begin{eqnarray}
\langle  \Psi_{0} | {\bf J}^{\rm }|~\Psi_{0} \rangle
&=&\langle \Phi_{0} | {\bf J}^{\rm }|\Phi_{0} \rangle +
\nonumber \\
&& \sum_{n \neq 0}{1 \over{E_{0} - E_{n}}} ~\Big( \langle \Phi_{0} |
{\bf J}^{\rm }|\Phi_{n} \rangle \langle \Phi_{n}|V_{\rm EM}|\Phi_{0} \rangle
+ \langle \Phi_{0}|V_{\rm EM} | \Phi_{n} \rangle \langle \Phi_{n} |
{\bf J}^{\rm }|\Phi_{0} \rangle \Big)~,  \label{emcor} \
\end{eqnarray}
and similarly for the charge density elastic matrix element
$ \langle  \Psi_{0} | {\rho}^{\rm }|~\Psi_{0} \rangle $.
These formulae have a simple Feynman diagrammatic interpretation shown
in Fig. \ref{f1}.  It is important to remember that the above
${\cal{O}}(\alpha)$ corrections do {\it not} constitute a {\it complete}
set. There are other graphs, e.g. those shown in Fig. \ref{f2} that
contribute to the same order in $\alpha$, but are not included in Eq.
\ref{emcor}. Our class of corrections, however, is gauge-invariant and
therefore a physically sensible subset.

\section{Results}

The first and foremost question is: do we observe any change in the static
properties of the nucleon, i.e., in the zero momentum transfer values of the
form factors? Then, the second question is: how do the effective radii of the
nucleon change? Both questions will be answered separately in each case.

\subsection{Mechanical Corrections}

\subsubsection{Corrections to static nucleon moments}

It turns out that two, the charge and the axial, out of three static
couplings remain unchanged. In the case of the isoscalar ``electric"
form factor (convection part of the vector current) at zero momentum
transfer this is a straightforward consequence of baryon number
conservation.  The isovector electric form factor at zero momentum
transfer is unrenormalized as well, in agreement with the
Ademollo-Gatto theorem \cite{ag64}. The axial coupling constant also
remains unchanged, but there does not seem to be as deep a reason for that
as for the charge conservation. In other words, this quark model as it
stands, does {\it not} predict the existence of an additional isovector
or any isoscalar axial couplings due to the Coulomb interaction between
the quarks and/or different quark masses. The magnetic moments,
however,  {\it are} corrected.

The change of variables from the equal mass three-body Jacobi
coordinates Eqs.~(\ref{eqiia}-c) to the  unequal mass case
(Eq.~\ref{eqvi}) amounts to a shift of the center of mass (CM) of the
system. That shift provides the main source of mechanical corrections.
The change of the oscillator frequencies plays a secondary role (see
tables I and II). E.g., the anomalous magnetic moments are renormalized
due to the change in the quark and nucleon masses, via the definition
of the magnetic moments:
\begin{eqnarray}
\langle N \uparrow |{e G_{{\rm M},N}(0)  \over 2 M_{N}}
{\mbox{\boldmath$\Sigma$}} | N \uparrow  \rangle =
\sum_{i = 1}^{3} \langle N \uparrow |
{e_{i} \over 2m_{i}}{\mbox{\boldmath$\sigma$}}_{i} | N \uparrow  \rangle ,
\end{eqnarray}
where $e G_{{\rm M},N}(0) = e_{N} + e \kappa_{N}$. It is important to note that
the quark mass $m_i$ is now an operator in isospin space
$m_{i} = {\bar m} + \Delta m {1 \over 2} \mbox{\boldmath$\tau$}_{i}^{z}$. Then
we find
\begin{mathletters}
\begin{eqnarray}
1 + \kappa_{p} &=& {8 m_{d} + m_{u} \over {3 m_{\lambda, p}}}
= {1 \over 9} \left(1 + {2 m_{u} \over  m_{d}}\right)
\left(1 + {8 m_{d} \over  m_{u}}\right)  \\
\kappa_{n} &=& -\left({2 \over 3}\right)
{m_{d} + 2 m_{u} \over {m_{\lambda, n}}}
= - {2 \over 9} \left(1 + {2 m_{d} \over  m_{u}}\right)
\left(1 + {2 m_{u} \over  m_{d}}\right) . \
\end{eqnarray}
\end{mathletters}
Corrections to nucleon static electroweak moments are summarized in
Table I.
Since the magnetic moments are the only static nucleon properties that
do receive corrections in this model, we will separate them into
various parts according to the nomenclature defined in Sec. II:
\begin{mathletters}
\begin{eqnarray}
{}^{u+d}G_{M}(0)^{p+n} &=& {1 \over 9} \left[7 + {m_{u} \over  m_{d}}
+ {m_{d} \over  m_{u}}\right] \simeq 1.0 \\
{}^{u-d}G_{M}(0)^{p-n} &=& ~~~ \left[3 + {m_{u} \over  m_{d}}
+ {m_{d} \over  m_{u}}\right] \simeq  5.0 \\
{}^{u-d}G_{M}(0)^{p+n} &=&
{1 \over 3} \left[{m_{d} \over  m_{u}} - {m_{u} \over  m_{d}}\right]
\simeq 0.008  \label{gmibc}\\
{}^{u+d}G_{M}(0)^{p-n} &=&
{1 \over 3} \left[{m_{d} \over  m_{u}} - {m_{u} \over  m_{d}}\right]
\simeq 0.008  . \label{gmibd}
\end{eqnarray}
\end{mathletters}
We can immediately use the results from
Eqs.~(\ref{isovector}, \ref{isoscalar}) to find
\begin{mathletters}
\begin{eqnarray}
G_{M,p}^{\rm NC}(0) &=& {1 \over 2} \left[(1 - 4\sin^2\theta_W )
		G_{M,p}^{\rm EM}(0) - G_{M,n}^{\rm EM}(0) - S_{M}(0)
		-{}^{u+d}G_M (0)^{p-n} + {}^{u-d}G_M (0)^{p+n} \right]
\nonumber \\
&\simeq & {1 \over 2} \left[(1 - 4\sin^2\theta_W )
		G_{M,p}^{\rm EM}(0) - G_{M,n}^{\rm EM}(0) - S_{M}(0)
					\right]\label{gmnca} \\
G_{M,n}^{\rm NC}(0) &=& {1 \over 2} \left[(1 - 4\sin^2\theta_W )
		G_{M,n}^{\rm EM}(0) - G_{M,p}^{\rm EM}(0) - S_{M}(0)
		+ {}^{u - d}G_{M}(0)^{p + n} + {}^{u+d}G_M (0)^{p-n}
							\right]
\nonumber \\
&\simeq& {1 \over 2} \left[(1 - 4\sin^2\theta_W )
		G_{M,n}^{\rm EM}(0) - G_{M,p}^{\rm EM}(0) - S_{M}(0)
		+ 2 {}^{u - d}G_{M}(0)^{p + n}
							\right] ~~, \
\label{gmncb}
\end{eqnarray}
\end{mathletters}
where the second lines of these two formulas represent this model's results.
Using the values in Eq.~(\ref{gmibc}-d), the proton weak magnetic form
factor is essentially unaffected, while the neutron weak magnetic form
factor gets a small but nonzero isospin-breaking correction of
approximately $0.3\%$.
{\it We conclude that, in this model, the proton magnetic moment measurements
in elastic parity-violating electron-nucleon scattering {\it can} be
directly interpreted as nucleon strangeness content, whereas the
neutron ones are subject to the above small IB correction.} It is not clear,
however, to what degree the above cancellation of IB terms in $G_{M,p}^{\rm
NC}$ is model dependent.

\subsubsection{Corrections to nucleon radii}

The $q^2$ dependent isospin-breaking effects are present in {\it all}
of the form factors. A few words about their evaluation are in order.
The nucleon wave function does not factor into three coherent parts due to
isospin breaking: the isospin and the spatial wave functions are now coupled
(see Eq.~\ref{isowave}) and that has to be taken into account.
With proper book-keeping of quark flavors in the spatial wave functions, one
derives the following formulae:
\begin{mathletters}
\begin{eqnarray}
e G_{E,p}^{\rm E}(q^2 ) &=& 2 e_{u} \langle u \rangle_{p} +
e_{d} \langle d \rangle_{p} \label{ffa} \\
e G_{E,n}^{\rm E}(q^2 ) &=& 2 e_{d} \langle d \rangle_{n} +
e_{u} \langle u \rangle_{n}  \label{ffb} \\
{e  \over 2 M_p} G_{M,p}^{\rm EM}(q^2 ) &=& {e \over 18} \left[{8 \over m_{u}}
\langle u \rangle_{p} + {1 \over m_{d}} \langle d \rangle_{p}
\right] \label{ffc} \\
{e \over 2 M_n} G_{M,n}^{\rm EM}(q^2 ) &=& - {e \over 9} \left[{1 \over m_{u}}
\langle u \rangle_{n} + {2 \over m_{d}} \langle d \rangle_{n}
\right]  \label{ffd} \\
 F_{A,p}^{\rm NC}(q^2 ) &=& {1 \over 3} \left[4 \langle u \rangle_{p} +
\langle d \rangle_{p} \right] \label{ffe} \\
 F_{A,n}^{\rm NC}(q^2 ) &=& - {1 \over 3} \left[4 \langle d \rangle_{n} +
 \langle u \rangle_{n} \right]~. \label{fff} \
\end{eqnarray}
\end{mathletters}
Here $\langle u \rangle_{p}$ is defined as the Fourier transform of
the spatial wave function matrix element of a $u$ quark in the proton i.e.
$\langle u \rangle_{p} = \langle \exp i {\bf q} \cdot {\bf r}_u \rangle_{p}$,
etc.. They are evaluated as
\begin{mathletters}
\begin{eqnarray}
\langle u \rangle_{p} &=& \langle \exp i {\bf q} \cdot {\bf r}_u \rangle_{p} =
\exp \left[- {{\bf q}^{2} \over 8}\left( R_{\rho p}^{2} + 3
\left({m_d \over M_{p}}\right)^2 R_{\lambda p}^{2}\right) \right] \label{upa}
\\
\langle d \rangle_{p} &=& \langle \exp i {\bf q} \cdot {\bf r}_d \rangle_{p} =
\exp \left[- {3 {\bf q}^{2} \over 2}
\left({m_u \over M_{p}}\right)^2 R_{\lambda p}^{2} \right]  \label{dpb} \\
\langle u \rangle_{n} &=& \langle \exp i {\bf q} \cdot {\bf r}_u \rangle_{n} =
\exp \left[- {3 {\bf q}^{2} \over 2}
\left({m_d \over M_{n}}\right)^2 R_{\lambda n}^{2} \right]
\label{unc} \\
\langle d \rangle_{n} &=& \langle \exp i {\bf q} \cdot {\bf r}_d \rangle_{n} =
\exp \left[- {{\bf q}^{2} \over 8}\left( R_{\rho n}^{2} + 3
\left({m_u \over M_{n}}\right)^2 R_{\lambda n}^{2}\right) \right] \label{dnd}
{}~.\
\end{eqnarray}
\end{mathletters}
Upon inserting these into Eqs.~(\ref{ffa}-f) and expanding in Taylor
series, we find the results that are shown in Table 2. They all turn
out to be small ($ \leq 1 \%$), and hence are not likely to be
distinguishable from the experimental uncertainties in the ``known" EM
form factors form factors.

One case is of particular importance, though, because of the vanishing
of the strange and neutron electric form factors in the static limit:
the weak NC charge radius. Since the nucleon strange radius is then
the first term in the Taylor expansion of the nucleon strange form
factor, it is of immediate experimental interest and will be measured
in e.g. parity-violating (PV) elastic ${}^{4}He({\vec e}, e')$ and forward
angle $p({\vec e,e'})p$ experiments. It is hence important that we know the
relevant IB corrections. As with the magnetic moment in the previous
subsection, we break up the total correction to the nucleon EM charge
radius into its isoscalar and isovector components
\begin{mathletters}
\begin{eqnarray}
{}^{u+d}\langle r^{2}_{E} \rangle^{p} &=&
{}~~{3\over2} \left({m_u \over M_{p}}\right)^2 R_{\lambda p}^{2} +
{1 \over 4}\left( R_{\rho p}^{2} + 3
\left({m_d \over M_{p}}\right)^2 R_{\lambda p}^{2}\right)
\label{vrp} \\
{}^{u-d}\langle r^{2}_{E} \rangle^{p} &=&
-{ 9\over2} \left({m_u \over M_{p}}\right)^2 R_{\lambda p}^{2} +
{3 \over 4}\left( R_{\rho p}^{2} + 3
\left({m_d \over M_{p}}\right)^2 R_{\lambda p}^{2}\right)
\label{srp} \\
{}^{u+d}\langle r^{2}_{E} \rangle^{n} &=&
{}~~{3\over2} \left({m_d \over M_{n}}\right)^2 R_{\lambda n}^{2} +
{1 \over 4}\left( R_{\rho n}^{2} + 3
\left({m_u \over M_{n}}\right)^2 R_{\lambda n}^{2}\right)
\label{vrn} \\
{}^{u-d}\langle r^{2}_{E} \rangle^{n} &=&
{}~~{9\over2} \left({m_d \over M_{n}}\right)^2 R_{\lambda n}^{2} -
{3 \over 4}\left( R_{\rho n}^{2} + 3
\left({m_u \over M_{n}}\right)^2 R_{\lambda n}^{2}\right)
\label{srn}~. \
\end{eqnarray}
\end{mathletters}
(In the isospin symmetric limit, e.g.,
$\langle r^{2}_{E} \rangle^{p} =
{}^{u+d}\langle r^{2}_{E} \rangle^{p} +
{}^{u-d}\langle r^{2}_{E} \rangle^{p} \rightarrow R^2$.)
Inserting the numerical values into these formulae gives
\begin{mathletters}
\begin{eqnarray}
{}^{u+d}\langle r^{2}_{E} \rangle^{p+n} / R^{2} - 1 &\simeq&  0.0 \%
\label{radiba}\\
{}^{u-d}\langle r^{2}_{E} \rangle^{p-n} / R^{2} - 1 &\simeq&  0.0 \% \\
{}^{u-d}\langle r^{2}_{E} \rangle^{p+n} / R^{2} &\simeq&  2.1 \%
\label{radibc}\\
{}^{u+d}\langle r^{2}_{E} \rangle^{p-n} / R^{2} &\simeq&  0.1 \% ~~. \
\label{radibd}
\end{eqnarray}
\end{mathletters}

There is a simple physical picture which roughly explains these
numbers.  The bulk of the effect arises from the shift in the position
of the center of mass when the constituent quarks are given different
masses.
In the case of the EM charge radius, we are evaluating a
weighted sum of the quark positions squared.
The isovector rms charge radius is weighted by the sign
of the quark charges, and thus in the neutron, although the shift of
the center of mass is in the opposite direction from that of the
proton, the sign of the quark charges is also switched, resulting in
the {\it same} increase in radius for proton and neutron. For the same
reason the isoscalar rms charge radius is very small.  Thus the bulk
effect is primarily due to ${}^{u-d}\langle r^{2}_{E} \rangle^{p+n}$.
This simple geometrical picture yields $\Delta \langle r\rangle^2/R^2
\approx 2 (\Delta m_Q /3 m_Q) \approx .8\%$ for proton and neutron
separately, and these are then {\it additive}. The remainder arises
primarily from the modification of oscillator frequencies due to
reduced mass effects, see Eq.~(\ref{masses}). We see that the complete
IB in the rms charge radius is due to fairly simple kinematical
effects, which, however, can only be evaluated if the CM motion is
correctly accounted for.

When we include strange form factors, we find
\begin{mathletters}
\begin{eqnarray}
\langle r^2_{\rm E} \rangle^{\rm NC}_{p} &=& {1 \over 2}
\left[(1 - 4\sin^2\theta_W )
		\langle r^2_{\rm E} \rangle^{\rm EM}_{p} -
			\langle r^2_{\rm E} \rangle^{\rm EM}_{n} -
					\langle r^2_{\rm E} \rangle_{s}
		-{}^{u+d}\langle r^2_{\rm E} \rangle ^{p-n} +
		 {}^{u-d}\langle r^2_{\rm E} \rangle ^{p+n}
			\right], \label{rnca}\\
\langle r^2_{\rm E} \rangle^{\rm NC}_{n} &=&
		{1 \over 2} \left[(1 - 4\sin^2\theta_W )
		\langle r^2_{\rm E} \rangle^{\rm EM}_{n} -
		\langle r^2_{\rm E} \rangle^{\rm EM}_{p} -
		\langle r^2_{\rm E} \rangle_{s}
		+  {}^{u - d}\langle r^2_{\rm E} \rangle^{p + n} +
                {}^{u+d}\langle r^2_{\rm E} \rangle ^{p-n}
							\right] ~~. \
\label{rncb}
\end{eqnarray}
\end{mathletters}
Using the numbers from Eqs.~(\ref{radiba} through d) and Table 2, the IB
contribution is apparently not negligible
in either proton or neutron weak radii, and is
indistinguishable from the strangeness radius contribution.
The fact that both
of the leading terms in Eq.~(\ref{rnca}) are suppressed, i.e. that the weak
neutral charge radius of the proton is naturally small, makes the
forward angle proton asymmetry an attractive place to look for
strangeness content.
However, the relatively large size of the isospin
breaking correction in this model,
(${}^{u-d}\langle r^{2}_{E} \rangle^{p+n}$ is more than 2\% of the
electromagnetic charge radius)
shows that isospin corrections may prove to be a
nontrivial effect which should be carefully taken into account.
The shift in our calculated value of $\langle r^2_E\rangle^{\rm NC}$
due to mechanical IB is approximately $+0.5$\% $R^2$ for both proton
and neutron.  For the proton, this shift is in fact
$+13$\% of the (very small) uncorrected Born approximation value
$.5 (1-4\sin^2\theta_W)$. Even if our model result is viewed merely as
an indication of the uncertainty introduced by isospin breaking, an
extraction of $\langle r^{2}_E\rangle_s$ by this means will be difficult
at or below a level of $\approx 2$\% of $R^2$.
This is equally relevant to the case of elastic ${}^{4}He({\vec e}, e')$
experiments, where the isoscalar radius is measured. Here again, the
isoscalar IB radius contributions add constructively (as would any
intrinsic nucleon strangeness radii, of course).
Note that the IB corrections due to the nuclear structure of
${}^{4}He$, i.e.  without the intrinsic {\it nucleon} IB breaking,
were recently calculated in Ref. \cite{don94} and were found to
be comparatively small.

\subsection{Electromagnetic Corrections}

As the first step in this analysis we take the lowest-lying baryon
resonances with all the quantum numbers equal to those of the nucleon,
but with ``wrong" isospin as the leading source of isospin admixtures.
A number of such resonances can be found in e.g. Table 1.1 of
Ref.\cite{bhad88} all at the ${\cal N} = 2$ level in the harmonic
oscillator model. Since the Coulomb potential does not carry orbital or
spin angular momentum, each of these is separately conserved. That
immediately eliminates all $L = 2$ states from this consideration. We
are left with only one multiplet of $L = 0$ resonances, the
$\Delta(1550 {\rm MeV})$,  with the required properties\footnote{In fact,
this resonance does not appear in the latest Particle Data Group tables
\cite{pdg94}, as it has been observed only by a single group
\cite{lon77}.}.

The mass of $\Delta (1550)$ is anomalously low, i.e., at about the same mass
as the ${\cal N} = 1$ negative-parity resonances.  Irrespective of its
present poorly confirmed experimental status, this is the only ${\cal
N} = 2$, SU(6) quark model state with the necessary quantum numbers;
hence we will use it here. Its isospin-symmetric wave function is
\begin{mathletters}
\begin{eqnarray}
^{2}10:~~|\Delta (1550) \rangle &=&
	{1 \over \sqrt{2}}  \left[ \chi^{\rho} \psi_{20}^{\rho} +
			\chi^{\lambda}\psi_{20}^{\lambda} \right] \phi_{S} \\
\psi_{20}^{\lambda} &=& {m \omega \over{\sqrt 3}}
		\left({m \omega \over{\pi}}\right)^{3/2}
	\left[{ \mbox{\boldmath$\rho$}}^{2} -
	        {\mbox{\boldmath$\lambda$}}^{2} \right]
					\exp\left(- {1 \over 2R^{2}}
		({ \mbox{\boldmath$\rho$}}^{2} +
		 { \mbox{\boldmath$\lambda$}}^{2}) \right) \\
\psi_{20}^{\rho} &=&
		{m \omega \over{\sqrt 3}}
		\left({m \omega \over{\pi}}\right)^{3/2}
	2 ({ \mbox{\boldmath$\rho$}} \cdot {\mbox{\boldmath$\lambda$}} )
					\exp\left(- {1 \over 2R^{2}}
		({ \mbox{\boldmath$\rho$}}^{2} +
		 { \mbox{\boldmath$\lambda$}}^{2}) \right) \\
\phi_{S}^{\Delta +} &=& {1 \over{\sqrt 3}}
		\left(uud + udu + duu \right) \\
\phi_{S}^{\Delta 0} &=& {1 \over{\sqrt 3}}
		\left(udd + dud + ddu \right) \
\end{eqnarray}
\end{mathletters}
The next admixed terms occur at the ${\cal N} = 4$ level, which ought
to be suppressed due to the
twice-as-large energy denominator. After some straightforward algebra
we find the admixture coefficients from Eq.~(\ref{eps}) are
\begin{mathletters}
\begin{eqnarray}
{\bf \varepsilon}_{p} &=& {- \alpha \over{3 R \Delta E \sqrt{3 \pi}}} =
4.1 \times 10^{-4} \label{eq22a} \\
{\bf \varepsilon}_{n} &=& - {1 \over{2}} {\bf \varepsilon}_{p} = - 2.0
\times 10^{-4} ~, \label{eq22b} \
\end{eqnarray}
\end{mathletters}
where $\Delta E = E_{N} - E_{\Delta(1550)} =  -612~ {\rm MeV}$ is the
experimental mass difference.

At this stage we must say a few words about the choice of free
parameters in this model. The constituent quark mass is fixed by
fitting the nucleon magnetic moments, and is then entirely consistent
with the mass of the nucleon.  The oscillator frequency can be fixed in
several ways, the two best known ones being:  (a) using the observed
spectrum of the nucleon resonances \cite{ik79}, or (b) using the
observed nucleon EM radius \cite{bhad88,yopr88}. The former leads to
reasonable values of $\omega$ only in the full-blown version of the
model with strong hyperfine interaction and it underestimates the
charge radius of the nucleon. The latter underestimates the energy gap
between the ground and the excited states. While this discrepancy may
be a serious cause of concern in attempts to make this model as
realistic as possible, it is only mildly worrying in our case.  Here,
we are interested in a first estimate of isospin-breaking effects on
electroweak form factors.  We have therefore taken the former approach,
and set $\omega \simeq 300 {\rm MeV}$, which leads to $\langle r^{2}
\rangle = R^2 = (0.62\, {\rm fm})^2$, an underestimate of the
experiment by about 30\%.  This does correctly yield the observed
$\Delta(1550) - N$ mass difference, however.  As one can see from
Eqs.~(\ref{eq22a}-b), an attempt to fix this up would somewhat {\em
decrease} the admixture coefficients.  In this sense, we are taking a
conservative point of view, and our results might be viewed as a rough
upper limit of the EM-induced isospin mixing effects.

The appropriate EM transition matrix elements are all isoscalar, i.e.,
they are identical for the charged and neutral members of the multiplet
\begin{mathletters}
\begin{eqnarray}
\langle \Delta |\rho ({\bf q})|~N \rangle
&=& \langle \Delta (1550) | \sum_{i = 1}^{3} e_{i}
\exp(i {\bf q} \cdot {\bf r}_{i}) |~N \rangle \nonumber \\
&=& -{e \over 12}\sqrt{2 \over 3} \left({{\bf q}^{2} \over{m \omega}}\right)
\exp \left({- {\bf q}^{2} \over{6 m \omega}}\right)  \\
\langle \Delta \uparrow | {e \over{2 M_{N}}}
{\mbox{\boldmath$\Sigma$}}^{\rm z} G_{\rm M}({\bf q})|~N \uparrow \rangle
&=&
\langle \Delta (1550) \uparrow | \sum_{i = 1}^{3} {1 \over{2 m_{i}}}
{\mbox{\boldmath$\sigma$}}_{i}^{\rm z} e_{i} \exp(i {\bf q} \cdot {\bf r}_{i})
|~N \uparrow \rangle  \nonumber \\
&=& { e \over 72 m}\sqrt{2 \over 3} \left({{\bf q}^{2} \over{m \omega}}\right)
\exp \left({- {\bf q}^{2} \over{6 m \omega}}\right)  \\
\langle \Delta \uparrow |{\bf J}^{(a = 3), {\rm z}}_{\rm A}
({\bf q})|~N \uparrow \rangle
&=& \langle \Delta (1550) \uparrow | \sum_{i = 1}^{3}
{1 \over 2} {\mbox{\boldmath$\sigma$}}_{i}^{\rm z} {
\mbox{\boldmath$\tau$}}_{i}^{a = {\rm z}}
\exp(i {\bf q} \cdot {\bf r}_{i})
|~N \uparrow \rangle  \nonumber \\
&=& {1 \over 36}\sqrt{2 \over 3} \left({{\bf q}^{2} \over{m \omega}}\right)
\exp \left({- {\bf q}^{2} \over{6 m \omega}}\right)
{}~.\
\end{eqnarray}
\end{mathletters}
None of these matrix elements survive the taking of the threshold
limit. Hence, these corrections do $not$ renormalize the static nucleon
electroweak couplings. The only exception is the neutron anomalous
magnetic moment which is renormalized by a small neutron EM mass shift
(see below). The corrections to the effective radii are shown in Table
III.  The main conclusion of this section is that all of the explicit
electromagnetic corrections due to admixtures with the $\Delta(1550)$
are extremely small, due to the smallness of $\alpha$, coupled with the
relatively small nuclear state overlaps, and relatively large energy
denominator.

In order to assess the reliability of our model, and that of our other
assumptions, we apply them to a well-understood case of isospin
breaking in nucleons:
most notably the nucleon mass difference $\Delta M_{N} \equiv M_{p} - M_{n}
\simeq - 1.3\ {\rm MeV}$.  We calculate this mass difference with all
free parameters in the model already fixed, compare it with experiment
and then discuss other possible approaches (schemes) to fixing the free
parameters.  We find \begin{equation} \Delta M_{N} = m_{u} - m_{d} +
\delta E_{C} \end{equation} where $\delta E_{C} \equiv E_{C}^{p} -
E_{C}^{n} = {\alpha \over 3} \langle {1 \over{|{\bf r}_{1} - {\bf
r}_{2}|}} \rangle $ is the difference of the Coulomb energies of the
proton and the neutron.

Using our Gaussian spatial wave functions, with $R = 0.62 {\rm\ fm}$,
the Coulomb energy $\delta E_{C} = {\alpha \over 3R} \sqrt{2 \over \pi}
\simeq  0.6 {\rm \ MeV}$.  This reproduces the correct (negative) sign
and order of magnitude (a few MeV) of the mass difference, but
quantitatively our computed value of -3 MeV is more than two times too
large.  Our assumptions are also well known to lead to other
unrealistic features of this model, such as Gaussian form factors. One
can in principle increase the small Coulomb energy by turning on the
strong hyperfine interaction between quarks, \cite{jmr92} but that
would take us far outside of the intended scope of this article.  We
could also just reverse the procedure and {\em fix} the constituent
quark mass difference from this calculation and the observed nucleon
mass difference, but that would open new questions with regard to the
applicability of those quark masses in the light meson sector. As far
as the IB effects in the nucleon are concerned, that procedure would
decrease their size as compared with our ``direct" procedure. In that
sense,  as discussed earlier, we are again taking a conservative point
of view.  The above discussion ought to be an indication of the
expected range of validity of this model: we may expect to have
calculated the correct sign and order of magnitude of the effects of
interest.  We leave the investigation of next-order effects such as
strong hyperfine interactions, and more realistic nucleon models as a
task for the future.

\subsection{Discussion}

We would like to make a few comments with regard to the place of this
investigation in the big picture of all ``radiative corrections" that
have to be evaluated. Our corrections form only one separately
gauge-invariant class of purely EM radiative corrections. We have
completely ignored EM quark vertex corrections and the associated
self-energies and Bremsstrahlung effects. These form another separately
gauge-invariant class of corrections that must be dealt with. If one
assumes elementary i.e. Dirac constituent quarks, then one sub-part of
such corrections, the anomalous magnetic moment of the quarks, are
finite and gauge-invariant by themselves and are given by Schwinger's
QED formula. This correction is manifestly of the same order of
magnitude as the above-found IB nucleon-structure-induced effects.  It
is not clear to us whether they, or indeed any of the additional
classes of graphs we have neglected, ought to be treated as intrinsic
IB effects, or as pure radiative corrections. Furthermore, even if we
were to include them into the former, the calculation would not be as
straightforward as in free space, due to the strong interaction
(``binding") effects.  Similar considerations hold for the EM
corrections to the axial current vertex.

On the formal side, the present model is subject to the criticism that
chiral symmetry is absent from it. If we assume that confinement and
chiral symmetry breaking are two essentially independent phenomena, in
accord with lattice QCD, then the situation can be remedied. The chiral
symmetry and its spontaneous breakdown, which endows constituent quarks
with their mass, can be installed into our model following e.g. Ref.
\cite{yopr88}, or Ref. \cite{vw91}. These are, of course, just two of many
chirally symmetric models for the constituent quark sector.
Such chiral
models allow a consistent evaluation of new kinds of diagrams such as
the EM correction to the constituent quark axial current vertex shown
in Fig. \ref{f2}. Diagrams of this sort create a constituent quark isoscalar
axial current coupling constant $g_{A, Q}^{S}$, which leads to an
isoscalar nucleon axial current coupling constant $g_{A, N}^{S}$
according to
\begin{eqnarray}
{g_{A, N}^{S} \over 2} = \langle N \uparrow |{g_{A, N}^{S} \over 2}
{\mbox{\boldmath$\Sigma$}}^{\rm z} | N \uparrow  \rangle =
\sum_{i = 1}^{3} \langle N \uparrow | {g_{A, Q}^{S} \over 2}
{\mbox{\boldmath$\sigma$}}_{i}^{\rm z}  | N \uparrow  \rangle
= {g_{A, Q}^{S} \over 2}~~. \
\end{eqnarray}
Manifestly, the result will depend on the model of the
constituent quarks that one adopts. Should the vanishing of this class of IB
corrections persist even at that next level of approximation, then we would
be allowed the straightforward interpretation of a nonzero isoscalar nucleon
axial coupling result, e.g. that of Ref. \cite{bnl92}
\footnote{ There are no
odd parity admixture corrections to leading order in the experiment
\cite{bnl92}, since that is a neutrino scattering experiment.},
in terms of strange quark contributions.

\section{Conclusions}

In summary, we have evaluated the IB admixture effects on the nucleon
electroweak form factors in the nonrelativistic constituent quark model.
We used the harmonic oscillator Hamiltonian as the confining interaction.
The results were classified according to a model-independent formalism
developed in Sect. II.

If we are ultimately interested in either extracting strange nucleon
form factors from electroweak data, or performing precision moderate
energy Standard Model tests,  we must first eliminate corrections such
as the isospin breaking effects discussed here. These are certainly
more ``conventional'' than the strangeness content, although also
interesting and potentially important in their own right. A direct
experimental measure of them, via e.g.  Eq.~(\ref{isovector}),
would be clearly valuable as well. Strange quark form factor estimates
range considerably \cite{rev94}, but typically yield from a small
fraction of a percent up to $\ge$ 10\% corrections to electroweak
structure. {\it A priori}, IB effects also modify electroweak structure
of the nucleon, but only at a level of $\cal{O}$($\alpha$). The issue
we have addressed here is effectively one of finding the coefficient in
front of $\alpha$ in these corrections. If one based one's judgement solely
on the observed nucleon mass splitting, then one would naively expect the
correction to be very {\it small}, i.e.  ${\cal O}(\Delta
M_N/M_N)\approx 0.1\%$, and {\em not} 1\%.

Our results show that the u-d mass difference can yield noticeably
larger modifications to electroweak form factors than naively expected
from the nucleon mass splitting. Nevertheless, most of the corrections,
with one possible exception, appear to be safely below the levels of
what might be considered ``interesting" strangeness content.  Indeed,
our results agree with the original expectation of $\cal{O}(\alpha)$
corrections, as well as with our assumption that the EM-mechanical
interference (cross-) terms may be neglected.  In the isoscalar charge
radius, we find corrections of several percent.  The main source of
this IB is simply the shift of the nucleon's center of mass. It is in
this regard that the present model is very well suited to the task at
hand, in contrast to e.g. the bag models.  We have, however, chosen the
model parameters to {\it maximize} these IB effects, so within the
subclass of IB corrections considered here,  our numbers should
represent an upper limit. E.g., choosing $\Delta m_Q$ to reproduce the
observed $M_p-M_n$ splitting in our model would halve the predicted
mechanical isospin corrections to the charge radius.  Since IB breaking
can, in principle,  interfere with the extraction of a relatively
modest strangeness-carrying nucleon electric form factor from
lepton-nucleon scattering experiments, our results indicate the need
for careful checking of these results in other nucleon models.

\acknowledgments

We would like to thank Volker Burkert for
helpful comments and Susan Gardner for emphasizing the importance of the
discrepancy between the two possible ways of fixing the oscillator
frequency in this model. This work was supported by the US DOE.
SJP acknowledges the support of a Sloan Foundation Fellowship.

\widetext
\begin{figure}
\caption{Feynman diagrams describing the EM isospin-breaking contribution to
the polar- and axial-vector currents. The solid line denotes the quarks, wavy
solid line is the photon and wiggly solid line is the neutral intermediate
vector boson $Z^0$. The shaded blob together with the three solid lines and
one double solid line leading to it denotes the nucleon wave function.}
\label{f1}
\end{figure}
\begin{figure}
\caption{An example of an ${\cal{O}}(\alpha)$ EM correction to the quark EM and
axial current
vertices of the type that may renormalize the nucleon magnetic moment and the
axial coupling constant. These diagrams are not evaluated in this paper.}
\label{f2}
\end{figure}

\begin{table}
\caption{The shift of nucleon static properties due only to mechanical IB
corrections in our model.
Here $M_{p} = 2 m_{u} + m_{d};~M_{n} = 2 m_{d} + m_{u}$ and
$x$ is defined by $x=\sin^2\theta_W$.
Numerical values used are given before Eq. \protect{\ref{qmasses}} in the
text.}
\begin{tabular}{llc}
 & {mechanical IB} & {relative correction} \\
 \tableline
{$G_{{\rm E}, p}^{\rm EM}(0)$} &
{$1$} & {$0$}
 \\
{$G_{{\rm E}, n}^{\rm EM}(0)$} &
{$0$} &
{$0$} \\
{$G_{{\rm M}, p}^{\rm EM}(0) $} &
{${1 \over 9} {M_{p} \over m_{d}} \left(1 + {8 m_{d} \over  m_{u}}\right)$} &
{$0.3 \%$} \\
{$G_{{\rm M}, n}^{\rm EM}(0) $} &
{$- {2 \over 9} {M_{n} \over m_{u}} \left(1 + {2 m_{u} \over  m_{d}}\right)$}&
{$< 0.01\%$}  \\
{$F^{\rm NC}_{{\rm A}, p}(0)$} &
{$+ 5 \over 3$} & {$0$}
 \\
{$F^{\rm NC}_{{\rm A}, n}(0)$} &
{$- {5 \over 3}$}&
{$0$}  \\
{$G_{{\rm M}, p}^{\rm NC}(0)$} &
{${1 \over 2} \left[(1 - 4 x)
		G_{M,p}^{\rm EM}(0) - G_{M,n}^{\rm EM}(0)
					\right]$} &
{$0.03\% $} \\
{$G_{{\rm M}, n}^{\rm NC}(0)$} &
{${1 \over 2} \left[(1 - 4 x)
G_{M,n}^{\rm EM}(0) - G_{M,p}^{\rm EM}(0) + {2 \over 3}
\left({m_{d} \over  m_{u}} -
{m_{u} \over  m_{d}}\right)	\right]$} &
{$-0.25\% $} \\
\end{tabular}
\end{table}

\begin{table}
\caption{Theoretical nucleon radii $\langle r^{2} \rangle$
with mechanical IB corrections,
as well as the relative change.  Notation is defined in the
text, after Eqs. (\protect{\ref{upa}} - d), except $x = \sin^2\theta_W $.
The isospin symmetric nucleon radius in each case is $R^2$ (except $\langle
r^2_{\rm E} \rangle^{\rm EM}_n = 0$) where we use $R = .62 {\rm\ fm}$. }
\begin{tabular}{llc}
 & {mechanical IB} &
{${\delta \langle r^{2} \rangle \over R^{2}}$} \\
 \tableline
{$\langle r^{2}_{\rm E} \rangle^{\rm EM}_{p}$} & {$R^{2}_{\rho ,p} + 3
\left({m_{d}^{2} -
m_{u}^{2} \over M_{p}^{2}}\right)R^{2}_{\lambda ,p}$}
& {~~1.1 \%} \\
{$\langle r^{2}_{\rm E} \rangle^{\rm EM}_{n}$} &
{$- {1 \over 2} R^{2}_{\rho ,n} + 3
\left({4m_{d}^{2} - m_{u}^{2} \over 2 M_{n}^{2}}\right)R^{2}_{\lambda ,n}$}
& {~~1.0 \%} \\
{$\langle r^{2}_{\rm M} \rangle^{\rm EM}_{p}$} &
{$\left({18 m_{u} m_{d} \over{8 m_{d} + m_{u}}} \right) \left[
{1 \over 3 m_{u}} \left(R^{2}_{\rho ,p} +
3\left({m_{d}\over M_{p}}\right)^{2} R^{2}_{\lambda ,p} \right) +
{1 \over 2 m_{d}}
\left({m_{u}\over M_{p}}\right)^{2}R^{2}_{\lambda ,p}\right]$}
& {~~0.4 \%} \\
{$\langle r^{2}_{\rm M} \rangle^{\rm EM}_{n}$} &
{$\left({6 m_{u} m_{d} \over{2 m_{u} + m_{d}}} \right) \left[
{1 \over 4 m_{d}} \left(R^{2}_{\rho ,n} +
3\left({m_{u}\over M_{n}}\right)^{2} R^{2}_{\lambda ,n} \right) +
{3 \over 2 m_{u}}
\left({m_{d}\over M_{n}}\right)^{2}R^{2}_{\lambda ,n}\right]$}
& {-0.1 \%}\\
{$\langle r^{2}_{\rm A} \rangle^{\rm NC}_{p}$} &
{${3 \over 5}\left[
R^{2}_{\rho ,p} + 3 R^{2}_{\lambda ,p} \left({m_{u}^{2} + m_{d}^{2}
\over M^{2}_{p}}\right) \right]$}
& {~~0.3 \% } \\
{$\langle r^{2}_{\rm A} \rangle^{\rm NC}_{n}$} &
{${3 \over 5}\left[R^{2}_{\rho ,n} +
3 R^{2}_{\lambda ,n} \left({m_{d}^{2} + m_{u}^{2}
\over M^{2}_{n}}\right)\right]$}
& {-0.3 \% }\\
\end{tabular}
\end{table}

\begin{table}
\caption{The shift of nucleon radii $\delta \langle r^{2} \rangle$ due to
purely electromagnetic
IB corrections, as calculated in section 4B.
Here ${\bf \varepsilon}_{p} = 4.1 \times 10^{-4}$;
${\bf \varepsilon}_{n} = - 1/2~ {\bf \varepsilon}_{p}$.}
\begin{tabular}{llllll}
 & {p} & {$~~~~{\delta \langle r^{2} \rangle \over R^{2}}$} & {n} &
{$~~~~~{\delta \langle r^{2} \rangle \over R^{2}}$} \\
 \tableline
{$\delta \langle r^{2}_{\rm E} \rangle^{\rm EM}$} &
{${{\bf \varepsilon}_{p} }\sqrt{2 \over 3}R^2$} &
{$~~3 \times 10^{-4}$} &
{${{\bf \varepsilon}_{n} }\sqrt{2 \over 3} R^2$} &
{$ - 2 \times 10^{-4}$} \\
{$\delta \langle r^{2}_{\rm M} \rangle^{\rm EM}$} &
{${ -{\bf \varepsilon}_{p} \over{3 }}\sqrt{2 \over 3}R^{2}$} &
{$-1 \times 10^{-4}$} &
{${ {\bf \varepsilon}_{n} \over{2 }}\sqrt{2 \over 3}R^2$} &
{$ - 8 \times 10^{-5}$} \\
{$\delta \langle r^{2}_{\rm A} \rangle^{\rm NC}$} &
{${-2 {\bf \varepsilon}_{p} \over{5 }}\sqrt{2 \over 3}R^2$} &
{$-1 \times 10^{-4}$} &
{${2 {\bf \varepsilon}_{n} \over{5}}\sqrt{2 \over 3}R^2$} &
{$ - 7\times 10^{-5}$} \\
\end{tabular}
\end{table}

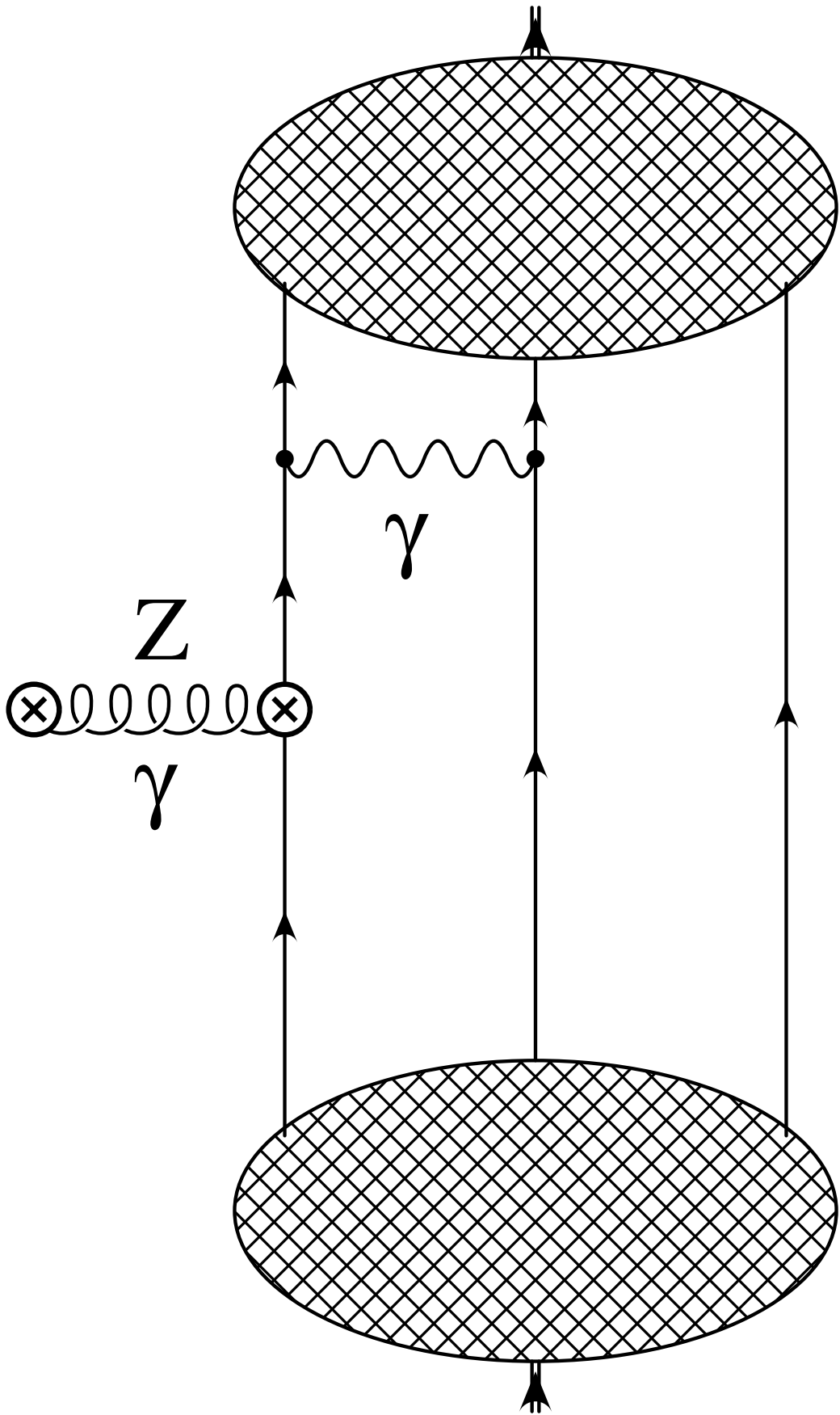
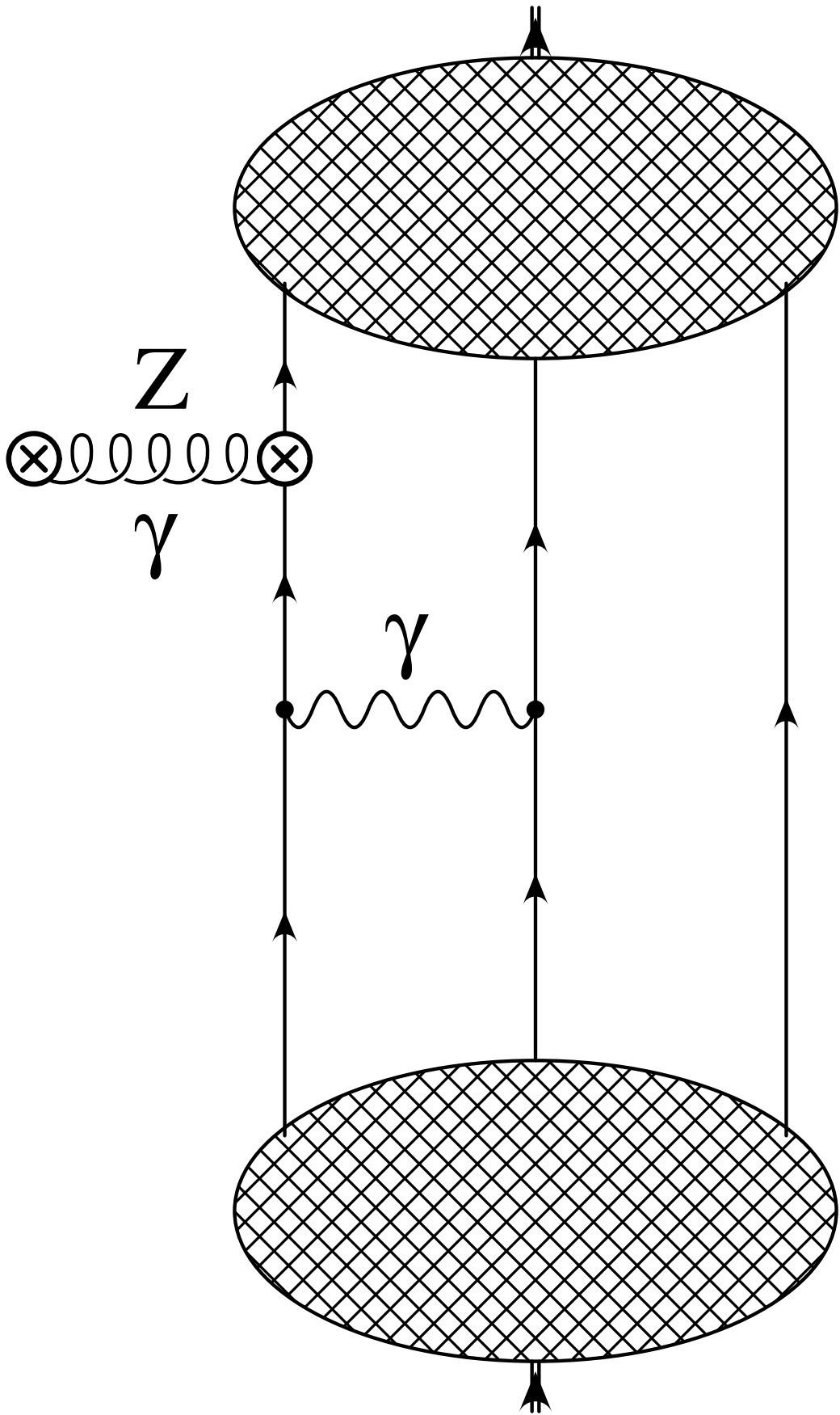
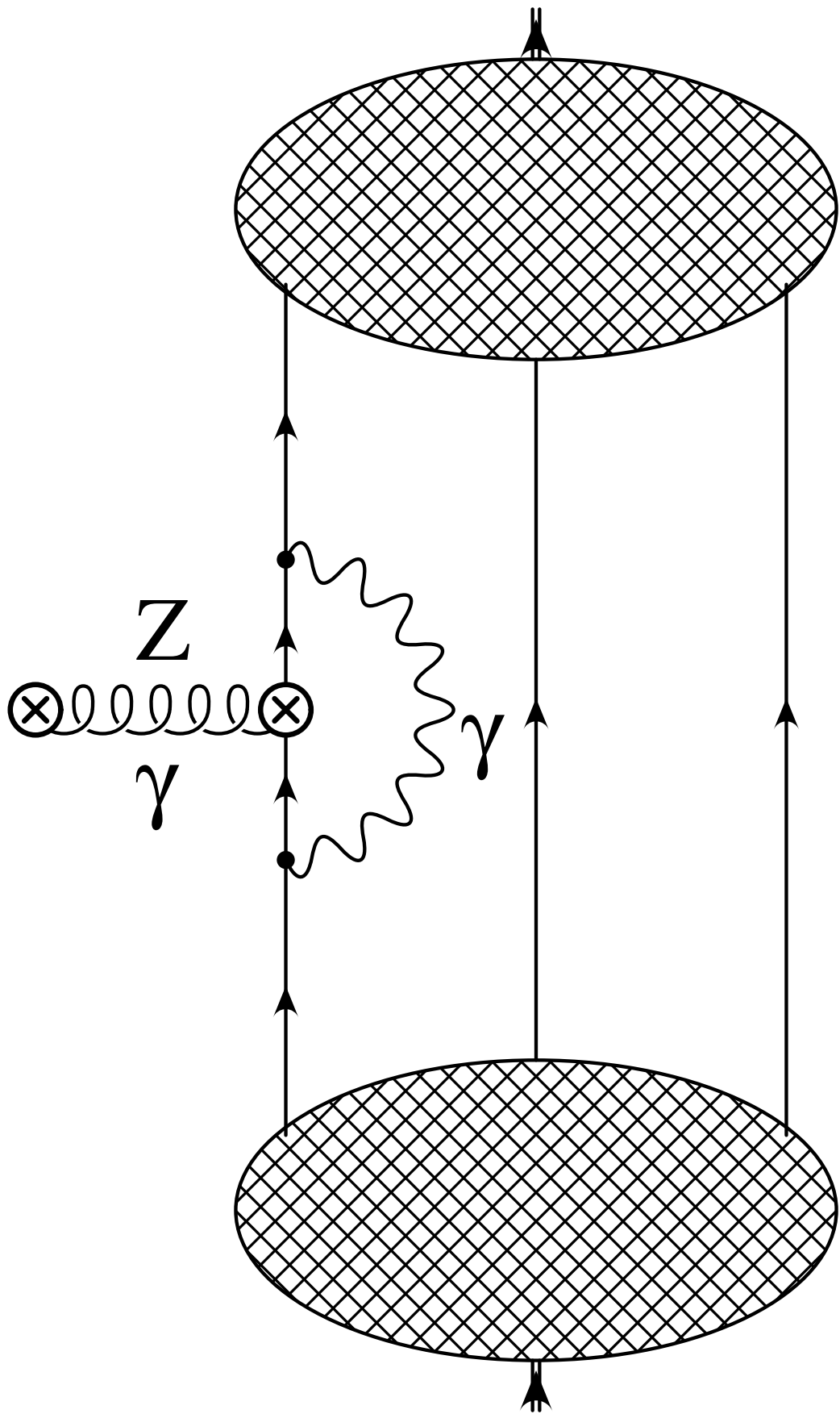
\end{document}